\documentclass[twoside,english]{elsarticle}
\usepackage[T1]{fontenc}
\usepackage[latin9]{inputenc}
\usepackage{graphicx}

\makeatletter

\providecommand{\tabularnewline}{\\}

\newenvironment{lyxlist}[1]
{\begin{list}{}
{\settowidth{\labelwidth}{#1}
 \setlength{\leftmargin}{\labelwidth}
 \addtolength{\leftmargin}{\labelsep}
 }}
{\end{list}}

\journal{arXiv}

\makeatother

\usepackage{babel}

\begin{document}

\title{Holographic Grid Cloud, a futurable high storage technology for the
next generation astronomical facilities\tnoteref{t1}}

\tnotetext[t1]{A feasibility study of a \textquotedblleft{}storage block\textquotedblright{}
for incoming astronomical implementations.}

\author{Stefano Gallozzi}

\ead{stefano.gallozzi@oa-roma.inaf.it / stefano.gallozzi@gmail.com }

\address[oar]{INAF-Osservatorio Astronomico di Roma, via Frascati 33, 00040 Monteporzio
Catone (RM) - Italy}
\begin{abstract}
In the immediate future holographic technology will be available to
store a very large amount of data in HVD (Holographic Versatile Disk)
devices. This technology make extensive use of the WORM (Write-Once-Read-Many)
paradigm: this means that such devices allow for a simultaneous and
parallel reading of millions of volumetric pixels (i.e. voxels). This
characteristic will make accessible wherever the acquired data from
a telescope (or satellite) in a quite-simultaneous way. 

With the support of this new technology the aim of this paper is to
identify the guidelines for the implementation of a distributed RAID
system, a sort of {}``storage block'' to distribute astronomical
data over different geographical sites acting as a single remote device.
To reach this goal it is needed a feasible platform (preferably open-source)
to facilitate authentication encryption and dissemination of web services
(including data storage), using indexed of appropriate hardware and
software (both physical and virtual). I identified the {}``eucalyptus''
open-source platform as best candidate for this purpose. Once developed
an appropriate IaaS (Infrastructure as a Service) architecture, it
is possible to virtualize required platforms (PaaS, Platforms as a
Service) and hence create services on-scientific-demand (SaaS, Softwares
as a Service). Any single service will be tied to the particular instrument
/ observatory and will be implemented at user's request through a
distributed sub-layer database of meta-datas intimately connected
to physical datas. From outside these services will be visible as
a part of a single resource: this is an effect of a main property
of distributed computing, the abstraction of resources. The end user
will only have to take care on connecting in a opportune and secure
mode (using personal certificates) to the remote device and will have
access to all (or part) of this potential technology. 

A Storage-Block+Services engineered on this platform will allow rapid
scalability of resources, creating a {}``network-distributed cloud''
of services for an instrument or a mission. It is recommended the
use of a dedicated grid-infrastructure within each single cloud to
enhance some critical tasks and to speed-up services working on the
redundant, encrypted and compressed scientific data. The power, the
accessibility, the degree of parallelism and of redundancy will only
depend on the number of distributed storage-blocks and on the potentiality
of the inter-network data connection: the higher this amount, the
greater will be throughput of the IT-system. A storage-block of this
kind is a meeting point between two technologies and two antithetical
computing paradigms: the Grid-Computing and Cloud-Computing. Extracting
the best from the two technologies will maximize benefits and minimize
disadvantages, thus allowing an optimal development of the infrastructure
for future observatories and demanding astronomical missions. 

In this paper I first present an overview of the technologies cited:
the Grid-Computing, the Cloud-Computing (i.e the Cloud-Storage), the
state of the art for Holographic data-storage and devices, after the
introduction I discuss the main issues on modern scientific data-storage
facilities. Finally I present a possible solution through the engineering
of a storage-block and a possible technical implementation for several
basic services on-demand.\end{abstract}
\maketitle

\section*{Introduction to IT-Technologies}

Let's start with a simple overview of the most used technologies cited
in this paper: the Distributed-Computing and in particular Grid-Computing
and Cloud-Computing (especially Cloud-Storage) focusing on similarities
and differences between these two paradigms. And the state of the
art for Holographic devices focusing on the possible improvements
for astronomical purposes.

\subsection*{Distributed-Computing}

A network which abstracts processing tasks can identify a Distributed-Computing
(DC) architecture. The abstraction in IT-technologies makes invisible
the actual complex processing underlying in a system and provides
a simple user-interface, with which users can interact easily. All
that is visible is the interface, which receives inputs and provides
outputs; how these outputs are computed is completely hidden to a
user.

It is possible to talk of DC when the computing elements of a network
are spread over large geographical area: in this context both Grid
and Cloud can be considered DC architectures. Distributing computation
over distant geographical locations makes also possible to spread
out the architecture costs accordingly.

There are great differences in the fundamental concepts of Grid and
Cloud Computing, but this does not imply that they are mutually exclusive:
it is possible to have a cloud within a computational grid, as it
is possible to have a computational grid as a part of a cloud. They
can also be the part of the same network, merely seen and represented
in two different ways.

\subsubsection*{Grid-Computing}

The Grid-Computing (GC) paradigm implies massive computer networks,
IT-assets and hence great investments. GC is an infrastructure which
relies on software (middle-ware) collecting together different hardware
resources to act as a single entity (a virtual whole). Such an entity
divides and directs pieces of programs on a large number of different
computers and processes single (or few) task (see {[}1{]}, {[}3{]},
{[}4{]}, {[}5{]}, {[}6{]}, {[}7{]}, {[}10{]}). Processed tasks are
always managed by a single primary computing machine (i.e. Computing
Element), making use of a batch queue system. The role of this machine
is to divide a single task in numerous sub-tasks, sending them to
different worker nodes and retrieving their outputs. A job's life-circle
is closed when all results are assembled in a single output, taking
care of monitoring the well-completion status of the whole work-flow
(see fig. 1). 

\begin{figure}
\begin{centering}
\includegraphics[width=8cm,height=8cm,keepaspectratio]{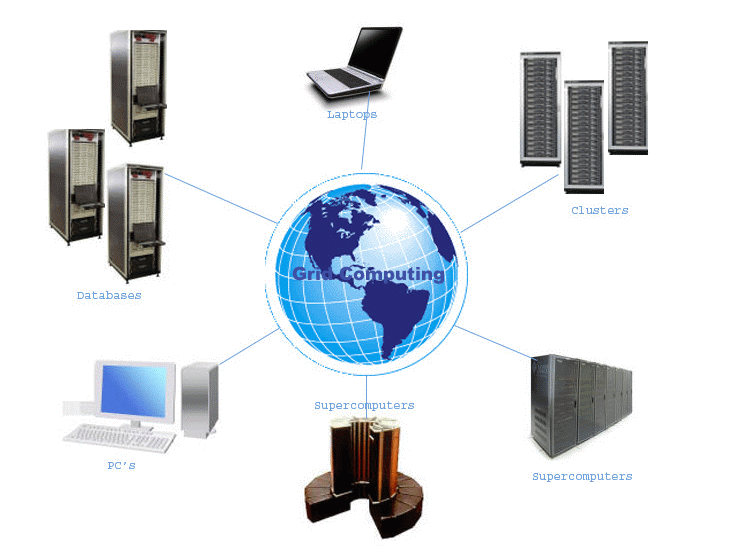}
\par\end{centering}

\caption{Grid-Computing: different distributed IT-resources are available to
a small number user as a unique whole to perform and run many jobs
efficiently.}

\end{figure}

GC offers two main advantages: 
\begin{enumerate}
\item it can easily make use of idle or unused processing power within distributed
computing (aggregation of processing power) 
\item the simultaneous execution of sub-independent task processes derived
by a parent job allows a significant reduction of a the total execution
time
\end{enumerate}
However GC has two main disadvantages: 
\begin{enumerate}
\item a job needs to be serialized in order to run in a data-grid network:
the code must be splitted in several tasks that are mutually exclusive
or independent. This is a different concept from that of parallel
computing where both the code and the hardware must be designed for
that purpose
\item the job I/O must have the same format under almost all circumstances
\end{enumerate}

\subsubsection*{Cloud-Computing}

The Cloud-Computing (CC) can be described as a shared access to computers
and their functionality via the Internet or a local network. This
is merely an extension of the object-oriented programming concept
of abstraction. 

The name CC refers to the fact that users don't know (or see) exactly
the underlying resources or their organization. A fraction of resources
(a resource-pool) is drawn by the {}``cloud'' when is needed and
returns to the {}``cloud'' when it is released (this concept is
called {}``on-demand cloud''). In other words, a {}``cloud'' is
a pool of resources (i.e. a set of machines and web-services) that
implements a CC system providing shared services (see {[}11{]}).

The key concept of {}``virtualization'' is widely used in the CC
paradigm: the ability to run {}``virtual machines'' (VM) on top
of a {}``hypervisor''. A VM is a software implementation of a machine
(computer with its own OS, kernel, library and applications), that
executes programs like a physical machine. The hypervisor provides
a unique and uniform abstraction layer for the underlying physical
machine. This means that by running VMs on the same hypervisor make
possible to match all user requests and needs for different services
(see fig. 2).

\begin{figure}
\begin{centering}
\includegraphics[width=8cm,height=8cm,keepaspectratio]{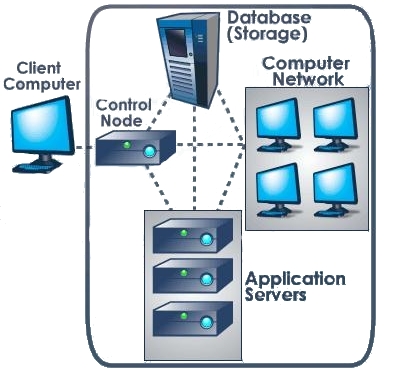}
\par\end{centering}

\caption{Cloud-Computing: different services available on-demand; the whole
cloud act as a single entity and the outside-client computer only
have to worry to authenticate to the cloud. }
\end{figure}

For the work purpose I will describe a particular CC system: the Cloud-Storage
(CS), where the primary goal of the CC architecture is the storage
of a huge amount of data (i.e. astronomical image and databases).
Following the advent of the new generation observatories, satellites
and instruments for astronomical investigation one of the main challenge
has been a fast, secure and affordable data storage. In last decade
the CS is experiencing a certain distribution because of the possibility
to save data off-site in storage systems maintained by a third party,
or, as I'm going to propose, build our own storage system. As any
CC service, the data storage is accessible from everywhere by simply
plugging-in the cloud as resource. In this way users do not have to
worry about redundancy, security or data-maintenance, but they only
have to authenticate themselves every time they need access to the
data.

CC offers several advantages: 
\begin{enumerate}
\item with CC it is possible to tap into multiple areas of expertise, using
a single resource: access to several servers an tasks through a single
access. In CC each server is specialized to perform well a single
task
\item CC provides organizations with massive scalability capabilities (also
out-sourcing of services on-demand, without the need of hardware/software
ownership)
\item in CC users may deploy its own resources (machines, storage, networks,
...) without the real need to buy all resources 
\end{enumerate}
However CC offers several disadvantages and concerns: 
\begin{enumerate}
\item in case of out-sourcing the asset is critically dependent on reliability
(and speed) of network: in the absence of a dedicated fast network
the CC use may become too expensive.
\item reliability of service because of the actual state of the www-network 
\item security of data may be guaranteed only with a combination of techniques:
a) encryption and the compression of data with latest encoding/compression
algorithms, b) the personal identification and authentication and
c) the authorization practices to access to all CC-resources.
\end{enumerate}

\subsubsection*{Grid-Computing vs Cloud-Computing}

As just noted, difference between GC and CC is hard to grasp because
they are not always mutually exclusive: both are used to economize
computing and maximize existing resources. Moreover the two architectures
use abstraction extensively and both have distinct elements which
interacts each other.

Anyway GC and CC are conceptually distinct; the main difference in
GC and CC lies in the way tasks are computed in the two environment: 
\begin{lyxlist}{00.00.0000}
\item [{GC}] in GC one large job is splitted in several sub-jobs and executed
simultaneously on multiple machines
\item [{CC}] in CC users can access to multiple services without the need
of great investments in the underlying architecture.
\end{lyxlist}
Concerning the end-user of the two DC approaches, the GC is designed
to run very large jobs for a small number of users, while the CC is
intended to support a large number of users. In this interpretation
CC involves selecting a particular provider and running in their data-center(s),
while GC involves a federation of multiple organizations (i.e. Virtual
Organizations, sometimes referred as Virtual Observatories).

\subsection*{Introduction to Holographic Data-Storage}

I analyzed different storage technologies to find the best suitable
device to be used in a future astronomical facility. The common magnetic
disks known as Hard Disk Device (HDD) are accessible, but because
of having moving parts their life time is too small to think at them
for future applications. The avoid of moving parts from the State
Solid Devices (SSD) rank them in a higher level of affordability;
they also make use of the volatile random-access memory (RAM) technology,
that permits a faster access to data, but the use of these devices
in a production grid presents other critical issues: a) the cost per
gigabytes is too high and 2) to store data permanently they need power
supply (i.e batteries). 

For these reasons I decided to migrate our attention to optical devices.

Taking a look on the technological development of optical devices
in later years (see table n.1), it is possible to remark the increasing
storage capability in the transition from one generation of technology
to another . 

\begin{table}
\begin{centering}
\begin{tabular}{|c|l|}
\hline 
Tech Generation  & Name, Description, Storage Capability (of a CDrom size disk)\tabularnewline
\hline
\hline 
1st & CDrom, storage $\sim700MB$ ($\sim1.5GB$ if doubleFaceCDrom)\tabularnewline
\hline 
2nd & DVD, storage $\sim4.7GB$ ($\sim9GB$ dualLayerDVD/doubleFaceDVD)\tabularnewline
\hline 
3rd & BD disks, storage $\sim25GB$ ($\sim50GB$ dualLayerBD/doubleFaceBD)\tabularnewline
\hline 
4th & 3D-Optical devices, storage $\sim1.0TB$ (see {[}15{]}, {[}16{]})\tabularnewline
\hline 
5th & Holographic HVD, storage $several\, TBs$ (see {[}9{]}, {[}14{]},
{[}17{]})\tabularnewline
\hline
\end{tabular}
\par\end{centering}

\caption{Here I present a summary on the main technological generation of optical
devices with improvements on storage capability.}

\end{table}

There are approximately the five generation of optical devices. While
common optical data storage media, such as CDs and DVDs (or BDs) store
data in a 2-D dimensional spiral track, which starts recording from
the center of the medium and extends until the disk boundary. The
data-recording and the data-reading is performed through the spiral
curve in a linear succession. The technical realization of such devices
is a set of reflective materials on an internal surface of a disk.
In order to increase storage capacity, it is possible to apply two
(or more) of these data layers, but their number is severely limited
since the addressing laser interacts with all the layers that it passes
through. To record (or read) the addressed layer. These interactions
cause noise that limits the technology to approximately 10 layers.

\begin{figure}
\begin{centering}
\includegraphics[width=8cm,height=8cm,keepaspectratio]{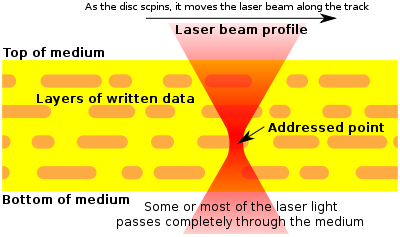}
\par\end{centering}

\caption{Schematic representation of a 3D optical storage disk section along
a data track, where the laser beam profile passes through the first
two layers and interacts only with the third layer since here the
light is at higher intensity.}
\end{figure}

It was possible to circumvent this issue by using an {}``addressing
methods'' in the 3D-optical medium. With this method only the specific
addressed voxel (i.e. volumetric pixel) interacts with the focused
laser-light, so the data-layers structure in a 3D optical disk write
(and read) bit-per-bit focusing only the addressed layer (and voxel),
while it passes through the other layers without interaction (see
fig. 3). In this context the operation of read/write data is necessarily
non linear as its technology is known as nonlinear optics. 

The last development of high-capacity data storage of optical devices
is the Holographic-Storage (HS) technology. The HS is an evolution
of the 3-D Optical Storage: its devices the Holographic Versatile
Disks (HVDs) has the same size of a DVD but differs from the 3D-optical
devices in the fact that it uses an holographic technology to store
(and read) data through all the volumetric surface. The technology
consists of two lasers, one green and one blue, which collimate in
a single beam. The green laser reads the coded data in the interference
fringes from a layer near the surface of the disk, while the blue
laser is used as reference beam to read the {}``servo-position''
that is an extra-information recorded in an aluminum layer near the
bottom of the disk. A dichroic mirror layer is placed between the
holograms and the aluminum layer so that the servo-information reflects
the green laser and transmits the blue laser, preventing interferences
(see fig. 4).

\begin{figure}
\begin{centering}
\includegraphics[width=8cm,height=8cm,keepaspectratio]{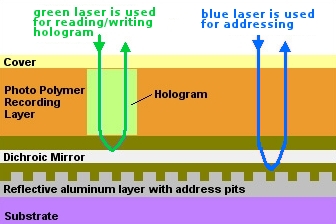}
\par\end{centering}

\caption{HVD schematic read/write structure of interaction: the blue laser
is used as a reference beam and takes care of addressing the data
read by the green laser.}

\end{figure}

At now the actual HVD transfer rate is $\sim1Gbit/s$ and it is planned
to achieve $\sim6TB$ of data storage for each disk in few years.
A part the high storage of HS technology, the real improvement of
the HVD devices is the WORM paradigm: Write Once Read Many times.
Unlike the traditional optical data storage, holographic storage is
capable of recording and reading millions of bits in parallel, enabling
data transfer rates much more great. Moreover the holographic technology
longevity, that is a grade of real technology reliability, is granted
for at least 50 years, that is much more of actual data storage systems
(RAID,NAS,HD,DVD and so on).

Like other optical media, HVD media is divided in Write-Once (where
the storage media undergoes irreversible changes) and Re-Writable
media that implements the physical phenomenon of the photo-refractive
effect in crystals (technical aspects are out of the target of this
papers and won't be discussed here, see {[}17{]} for further informations).

\section*{The Issue: huge data storage in astronomical facilities }

Demanding astronomical observatories (from the Earth and from the
space) equipped with last generation instruments, will produce very
large amount of data every day. In this scenario one of the most tricky
and challenging problem for the astronomical community is the efficient
storage of a such amount of data without information losses. Sometimes
technology is not affordable and sometimes engineers are not good
enough to ensure an infrastructure well built around a scientific
project. The analysis of very large amount of data is a fundamental
requirement of modern scientific research: for instance large area
surveys needs large amount of stored data and top level astronomical
and cosmological studies can not be performed without systematic and
accurate large programs.

In countless cases projects, missions and/or instruments produce so
many data that their storage facility are unable to dispose and store
efficiently. In extreme cases there are instrument forced to delete
intermediate files (sometimes sensible astronomical-data) to ensure
the storage-capacity for the principal target of the observation.
In these extreme cases it is excluded the possibility of later investigation
on the serendipitous field observed near the target of the observation.
Other instruments delete images and work only on catalogs, this excluding
the possibility to perform 2nd target level science with unused science
images. 

In great projects, where consortium of several institutions work together
to reach the same scientific goal, this critical issue is often faced
in the optical of a single-huge (even mirrored) repository/database,
where all partners can access to data and perform particular tasks.
These tasks may involve the computation of astronomical quantities
and send back the output of the computation to the central storage
facility. This approach is rarely efficient and critically depends
on the high efficiency of the network grid as well as the inter-connections
between parts.

As I'm going to expose, to overcome this disturbing problem new technologies
discussed in the previous sections may come into help .

\section*{The Solution: distributed+scalable Storage-Block for astronomical
databases }

The idea suggested in this paper is the possibility to identify one
coherent architecture/infrastructure that can be identified as standard
approach to those problems. The solution proposed is engineered to
be scalable: this property will allow the division of costs and encourage
the cooperation between various research projects and institutions. 

I identify the Cloud Computing paradigm as the base platform where
to build a distributed storage block (see fig. 5 for a graphical explanation
of the desired data-flow). 

When a particular instrument acquires a scientific image in a temporary
local repository, the bit code is sent to many indexed resources where
a parallel software takes care of split, encrypt and compress all
the image data in sub-packages and store them in a RAID distributed
repository.

The function of the RAID repository is the core of the {}``storage-block''.
Such a block will have a Grid-Computing like infrastructure primarily
composed by:
\begin{itemize}
\item a STORAGE ELEMENT (SE), which takes care of record redundantly all
sub-package data (i.e. where physically relies the RAID-block)
\item a COMPUTING ELEMENT (CE), where to queue users requests to physical
data. In the CE will relies a particular DATABASE composed of {}``meta-data
descriptors'' (i.e. informations about the data). The role of the
database is the information browsing over the encrypted sub-package
data without a direct physical access to them 
\item a WORKER ELEMENT (WE), where to decompress/decrypt sub-packages on
the fly accordingly to DATABASE descriptors. At this level the sub-packages
are link, decrypt and assembled together in order to make physical
data available to users (any physical data is a resource that is built
and destroyed by the user on-demand)
\end{itemize}
The division on elements is just representative of the necessary functions:
it is possible to build a storage-block which performs all described
functions within a single machine).

Since security is one of the most critical issue in the cloud computing,
it is necessary to encrypt the data according to personal key or certificates:
the access to the cloud resource is granted to a user through the
same authentication (UI+network).

\begin{figure}
\begin{centering}
\includegraphics[width=12cm,height=12cm,keepaspectratio]{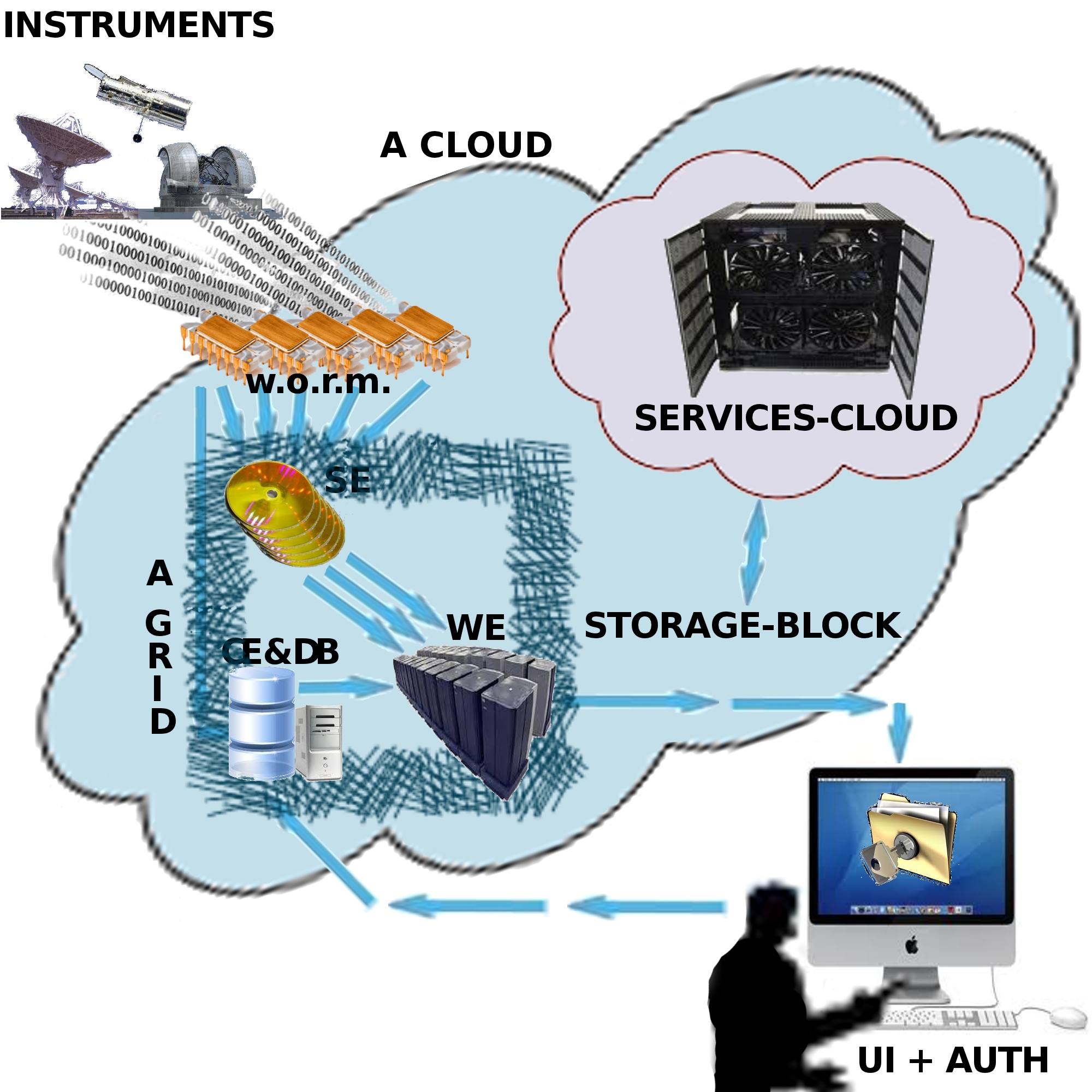}
\par\end{centering}

\caption{Schematic representation of the predicted work data-flow for the Storage-Block.
At higher level the instrument contact the cloud to fast store sub-packages
of data. The User Interface (UI) takes care of authenticate securely
the user to the cloud; the user demand (and construct) an instance
to the Computing Element (CE) that gives the meta-data information
about the physical data requested, the Working Element (WE) fast retrieve,
decrypt and assemble the data and make it accessible to the user outside.
The procedure appear invisible to the end-user that only request/dismiss
the remote resource. Moreover disseminated service-clouds may offer
to the end-user different services on the cloud stored-data.}
\end{figure}

The critical step is to ensure the most I/O velocity possible for
data transmission from SE $\rightarrow$ WE, that is the choice to
use the next generation Holographic Storage technology, where it is
possible to Write Once / Read Many. Using opportune parallel architecture
for the WE the composition of sub-package in final data time latency
will be minimized.

As well as an end-user access to the Cloud Computing, he will dispose
of the remote resource and a set of tools/services to be invoked on-demand
to the remote system. 

In this context it is possible to identify a Pipeline-Element block
to be part of the main Cloud (but not the Grid-part): such an element
will publish on the cloud several services including the possibility
of performing basic operation on private (and public) data persistent
in the Storage Block. The end user will only have to take care of
securely authenticate to the cloud, but will be totally obscured about
what's really happens in the computational sub-layers. Adding a service
element to the primary storage-cloud transforms the desired storage-block
in a storage+service-block and opens further perspectives on the IT
technologies applied to astronomical data.

As last item I remark the possibility to improve the power, the reliability
and the security of the whole Holographic Grid Cloud system just adding
scalable resources (i.e. storage-blocks). 

Other improvement can be the replication of the whole architecture
in different geographical areas (this adds degrees of redundancy),
but to distribute fairly the workload especially over the network
it is necessary to make all resources indexes in the cloud. 

Moreover the RAID redundancy must be distributed over geographical
areas so that a local failure will not corrupt the redundant data-array.
A key role in the data-reconstruction is played by the meta-data database
facility, that will collect all sub-packages information to easily
reconstruct all parent data. Any storage-block needs a database mirror
that must be continuously synchronized and committed with a NTP central
server for time synchronization.

\section*{A possible technical implementation}

To quantify the actual feasibility of the project it is necessary
to design a feasible technical implementation. This is mainly out
of the scope of this paper, but I can provide several guidelines to
overcome the main issues in order to build the desired Storage+Facility
Block for the next astronomical facility.

\subsection*{How to build a customized Cloud}

The Cloud-Computing is an abstract concept whose practical realization
is quite impossible without having recourse to a software suite that
performs the dirty work on our behalf. 

I identified {}``The Eucalyptus Open-source Cloud-computing System''
(see {[}7{]}, {[}8{]}, {[}11{]}, {[}12{]}) as the best suitable software-suite
for our purpose. With Eucalyptus it is possible to implement an IaaS
service (Infrastructure as a Service), which makes the IT-administrator
able to create (and control) virtual machines and instances deployed
across a variety of physical resources. Those resources may be linked
together and virtualized to obtain a small private and versatile cloud.
With Eucalyptus comes security, authentication, virtualization and
abstraction, so that a user can access to the system without any knowledge
of the inner mechanism.

\subsection*{How to build an inner Storage Holographic Element}

At present this is the most tricky issue because of the actual technological
evolution of such devices: there are only prototypes or high expensive
HVD burners/readers. But I am confident because the evolution of such
devices is evolving steadily, making this technology enough mature
in few years. 

However it is possible to implement a prototype with a simple BD dual-layer
juke-box (or a diffuse hard disk RAID), planning to migrate as the
holographic-technology will be enough stable and affordable to be
replaced the main storage unity. 

The challenge for this module is the development of an upper module/library
to ensure the splitting, encryption and compression of data-flow.
This library should have a split/assemble instance, an encode/decode
instance, a compress/decompress method, as well as the possibility
of communicate with the cloud authentication layer in order to obtain
a random crypt key to encode/decode data. 

This library can be assembled using different open-source softwares:
at now the GNU-parallel open source program (see {[}30{]} for further
insights) splits large files and execute parallel job on several CPUs
(even on remote nodes), the MD5/SSL crypts and decrypts image data
and the fpack/funpack (HEASOFT library, see {[}31{]}) that highly
compresses and decompresses fits images.

\subsection*{How to build an inner private Grid}

The Grid architecture used in the storage-block project is only private
grid, so it is not necessary to start-up an official Data-Grid node
with any test-bed connection. The private grid will assemble the hardware
resources and act as a single entity of inter-communication. The main
role of such a grid is the possibility of publishing on the parent-cloud
the user-queries for files (and eventually services). 

It is possible to implement and obtain by the grid middle-ware only
required services. A possible source for these services is the Globus
Toolkit (ref. {[}2{]}), but other packages listed in the Open Grid
Forum alliance (see {[}18{]}) have similar usability. 

The main role of such a grid is the speed-up of the data reconstruction:
using the synchronized meta-database accessible by the CE, provides
the allocation indexed of sub-data-packages and the information for
decryption. The working element will find in the cloud the user-authentication
information needed to assemble required data in a quickly and reliably
operation. 

A concrete possibility to speed-up the $CE\leftrightarrow CW$ inner-link/tasks
is to build a Working Element equipped with Graphics Processing Units
(GPUs) using the data-reconstruction as an optimized 3D-graphical
manipulation inside the Computing Element graphic device (see {[}32{]}
and {[}33{]} for references). The challenge for this approach is to
best fit-in the data-RAM requirements with the RAMDAC frame buffer
of the GPU.%
\footnote{An optimized version of the RAM for the computer system graphic devices.
RAMDAC stands for Random Access Memory Digital-to-Analog Converter,
which allows a higher clock frequency for the RAMDAC respect to the
RAM.%
} 

The best performances for the IT-system is reached when the hardware,
software and/or middleware configuration adopted minimizes the time-latency
for data requests and maximizes the throughput for the data transfer.
Several tests and experiments are expected to be needed in order to
reach this goal.

\section*{Conclusions}

In this work I present an affordable way to build a scalable storage
block for the next generation demanding astronomical facilities. 

The intent of the work is propose a futurable technology that is:
a) all-in one; b) highly efficient; c) almost scalable; d) highly
customizable; e) on scientific-demand.

This paper wants to be a starting point for a general discussion about
the storage-technology to adopt in next decade astronomical missions,
when a very hard I/O is expected. 

Excellent candidates to discuss this architecture can be the ESA-EUCLID
mission (see {[}25{]}, {[}26{]}, {[}27{]}, {[}28{]}, {[}29{]}) and
the Chernenkov Telescope Array (see {[}19{]}, {[}20{]}, {[}21{]},
{[}22{]}, {[}23{]}, {[}24{]}).

The exposed idea is to join several advantages from the two different
approaches: the Cloud-Computing and the Grid-Computing. This work
finds a meeting point where the two technologies gives the best. 

The last intriguing scenario proposed in this paper is the perspective
of using Holographic data storage facility to implement the WORM paradigm:
this possibility enhances the throughput for the whole IT-system. 

I have also proposed several technical guidelines to build a prototype
to be tested and proposed to the scientific community.

\section*{Acknowledgments}

I wish to thank all LBT/LSC and LBC team for the support and the ideas
given for the completion of this work, especially the A. Grazian,
M. Castellano and F. Pedichini efforts.

\section*{References}
\begin{lyxlist}{00.00.0000}
\item [{{[}1{]}}] The anatomy of the grid: Enabling scalable virtual organizations
- Foster, Kesselman, et al. - 2001 
\item [{{[}2{]}}] Globus: A metacomputing infrastructure toolkit - Foster,
Kesselman - 1997 
\item [{{[}3{]}}] The physiology of the Grid: an open grid services architecture
for distributed system integration - FOSTER, N 
\item [{{[}4{]}}] Xen and the art of virtualization - Barham, Dragovic,
et al. - proceedings SOSP 2003 
\item [{{[}5{]}}] The Grid: Blueprint for a New Computing Infrastructure
- Foster, Kesselman - Margan Kaufmann, 2003 
\item [{{[}6{]}}] Grid Computing Making Global Infrastructure a reality
- Berman, Fox, et al. - Wiley and Son, 2003 
\item [{{[}7{]}}] Virtualization for high-performance computing - Mergen,
Uhlig, et al. 
\item [{{[}8{]}}] The Eucalyptus Open-Source Cloud-Computing System - Nurmi,
D.; Wolski, R.; Grzegorczyk, C.; Obertelli, G.; Soman, S.; Youseff,
L.; Zagorodnov, D.; Comput. Sci. Dept., Univ. of California, Santa
Barbara, MD - 2009
\item [{{[}9{]}}] Optimizing holographic data storage using a fractional
Fourier transform - Nicolas C. Pégard and Jason W. Fleischer - Optics
Letters, Vol. 36, Issue 13, pp. 2551-2553 (2011)
\item [{{[}10{]}}] NSF Tera Grid Project. www.teragrid.org
\item [{{[}11{]}}] Elastic Compute Cloud, Amazon EC2, http://aws.amazon.com/ec2/
\item [{{[}12{]}}] Eucalyptus Public Cloud (EPC), http://eucalyptus.cs.ucsb.edu/wiki/EucalyptusPublicCloud/
\item [{{[}13{]}}] Programming the Grid: Distributed software components,
D. Gannon, 2002
\item [{{[}14{]}}] Holographic Data Storage, IBM Journal of Research and
development, 2008
\item [{{[}15{]}}] Three dimensional optical data storage using Photochromatic
Materials, S. Kawata and Y. Kawata, Chem Rev. 2000 100, 1777
\item [{{[}16{]}}] Three dimensional optical storage, G.W. Burr, SPIE Conference
(2003), paper 5255-16
\item [{{[}17{]}}] Photorefractive Organic Thin Films, Ed. Z. Sekkat and
W. Knoll, Elsevier USA, ISBN 0-12-635490-1
\item [{{[}18{]}}] Open Grid Forum, http://www.gridforum.org
\item [{{[}19{]}}] Cherenkow Telescope Array, CTA, http://www.cta-observatory.org
\item [{{[}20{]}}] Design Concepts for The Cherenkov Telescope Array, The
CTA Consortium, (2010), available at arXiv 1008.3703
\item [{{[}21{]}}] Teraelectronvolt Astronomy, J.A. Hinton \& W. Hofmann,
(2010), Ann. Rev. Astron. Astrophys., 47:523, or at arXiv 1006.5210
\item [{{[}22{]}}] High energy astrophysics with ground-based detectors,
F. Aharonian, J. Buckley, T. Kifune \& G. Sinnis, 2008, Rep. Prog.
Phys., 71:096901
\item [{{[}23{]}}] TeV Gamma-Ray Astronomy: The Story So Far, T. Weekes,
2008, AIP Conference Proceedings, 1085:3
\item [{{[}24{]}}] The Status and future of ground-based TeV gamma-ray
astronomy. A White Paper prepared for the Division of Astrophysics
of the American Physical Society\textquotedbl{}, J. Buckley et al.,
2008, available at arXiv:0810.0444
\item [{{[}25{]}}] ESA, EUCLID Mission, http://sci.esa.int/science-e/www/area/index.cfm?fareaid=102
\item [{{[}26{]}}] Euclid Mission Assessment Study - Executive Summary
(Thales Alenia Space, EADS Astrium), Thales and EADS Astrium, 2009
\item [{{[}27{]}}] Euclid definition study report (Red Book), ESA, ESA/SRE(2011)12
\item [{{[}28{]}}] Technical Review Report - Euclid, ESA, SRE-PA/2009/051
\item [{{[}29{]}}] Euclid assessment study report (SRE-2009-2), ESA, 2009
\item [{{[}30{]}}] GNU Parallel - The Command-Line Power Tool, O. Tange,
The USENIX Magazine, February 2011:42-47
\item [{{[}31{]}}] HEASoft - NASA's HEASARC Software, http://heasarc.nasa.gov/lheasoft
\item [{{[}32{]}}] Evolution of the Graphical Processing Unit, Thomas Scott
Crow, dr.Frederick C. Harris Jr, Master of Science University of Nevada,
2004
\item [{{[}33{]}}] GPUs - Graphics Processing Units, Minh Tri Do Dinh,
Vertiefungsseminar Architektur von Prozessoren, SS 2008
\end{lyxlist}

\end{document}